\documentclass[pra,twocolumn,a4paper,nofootinbib]{revtex4-1}

\usepackage{hyperref}
\usepackage{graphicx}
\usepackage{bm}
\usepackage{amsmath}
\usepackage{amssymb}
\usepackage{amsthm}

\begin{document}

\title{Revisiting consistency conditions for quantum states of systems on closed timelike curves:  an epistemic perspective}

\author{Joel J. Wallman}
\author{Stephen D. Bartlett}
\affiliation{School of Physics, The University of Sydney, Sydney, New South Wales 2006, Australia}

\date{13 May 2010}

\begin{abstract}
    There has been considerable recent interest in the consequences of closed timelike curves (CTCs) for the dynamics of quantum mechanical systems.  A vast majority of research into this area makes use of the dynamical equations developed by Deutsch, which were developed from a consistency condition that assumes that mixed quantum states uniquely describe the physical state of a system.  We criticize this choice of consistency condition from an epistemic perspective, i.e., a perspective in which the quantum state represents a state of knowledge about a system.  We demonstrate that directly applying Deutsch's condition when mixed states are treated as representing an observer's knowledge of a system can \textit{conceal} time travel paradoxes from the observer, rather than resolving them.  To shed further light on the appropriate dynamics for quantum systems traversing CTCs, we make use of a toy epistemic theory with a strictly classical ontology due to Spekkens and show that, in contrast to the results of Deutsch, many of the traditional paradoxical effects of time travel are present.
\end{abstract}


\maketitle

\section{Introduction}

While chronology violations (i.e., objects traveling backwards in time) have never been observed, they can occur in solutions of classical general relativity in the form of closed timelike curves (CTCs)~\cite{godel49, gott91, morris88, ori07, tipler74}. Without a quantum theory of gravity (and potentially, more empirical data on the initial conditions of the universe), the possibility of CTCs is unknown. Nevertheless, thought experiments that determine the potential implications of CTCs for quantum mechanics and general relativity could provide insight into a range of foundational questions. Specifically, as emphasized by Deutsch \cite{deutsch91}, CTCs are an extreme phenomenon that require additional assumptions in addition to standard non-relativistic quantum mechanics.  Different interpretations of quantum mechanics naturally lead to different additional assumptions, which in turn may lead to physically distinguishable predictions within the modified theory. We can therefore use thought-experiments based on quantum particles traversing CTCs to compare different interpretations of the theory.

A fundamental issue that interpretations of quantum mechanics seek to address is the relationship between the mathematical formalism of quantum mechanics and physical reality (if a physical reality is assumed to exist). One such relationship involves deciding whether quantum states are \textit{ontic} or \textit{epistemic} states. The ontic state of a system is the physical (i.e. objective) state that the system is in, whereas an epistemic state assigned to a system by an observer is an object that reflects the observer's knowledge about the system. An epistemic state is typically a probability distribution over the space of ontic states. From these definitions, one can see that the ontic state of a system is unique, while epistemic states in general are not (i.e., different observers can describe a system by different epistemic states).

The issue of ontic and epistemic states is particularly important when considering CTCs because, as we will argue, the interpretation of quantum states as either ontic or epistemic will naturally lead to different assumptions about how quantum systems behave in the presence of CTCs. For example, Deutsch~\cite{deutsch91} studied various time travel scenarios in a classical model and then in a quantum model motivated by an ontic interpretation of quantum states (specifically, the Everett interpretation~\cite{everett}). While in the classical model, paradoxes could occur, Deutsch argued that no paradoxes occur in his quantum treatment. Although all paradoxes are resolved, the resulting theory is not standard quantum theory, but a new nonlinear theory. Some of the implications of the deviation from quantum mechanics include: that quantum computers with access to CTCs can efficiently solve \textit{NP}-complete problems~\cite{bacon04}; that classical and quantum computers with access to CTCs have the same computational power~\cite{aaronson08}; that observers can perfectly distinguish nonorthogonal states (rendering quantum cryptography vulnerable to an adversary with access to CTCs)~\cite{brun08}; the evolution of chronology-respecting particles can be a discontinuous function of the initial state \cite{frey} and that information can be stored without degradation \cite{durham}. (For alternate perspectives on these results, see \cite{bennett, lloyd2010, dasilva2010, svetlichny2009}.) 

All of these results stem from the assumptions that Deutsch advocated, based on his interpretation. These assumptions, motivated by an ontic interpretation of quantum states, are particularly unconventional, in part because they require that \emph{mixed} quantum states are ontic. (Although it is common for pure quantum states to be interpreted as ontic, most interpretations view mixed states as epistemic, i.e., reflecting an observers lack of knowledge.)

In this paper, we demonstrate that using different interpretations of the quantum state naturally leads to different assumptions for quantum systems traversing CTCs than Deutsch's. Specifically, we argue that the behavior of epistemic states describing quantum systems traversing CTCs allows for paradoxes, just as in the classical case, and that often the theory will \emph{conceal} paradoxes rather than resolving them. We also examine how a toy epistemic theory with a classical ontology behaves in the presence of CTCs in order to gain some insight into what the existence of CTCs would imply for an epistemic interpretation of quantum mechanics.

The paper is structured as follows. In Sec.~\ref{sec:paradoxes} we discuss the concept of time travel paradoxes and fix our terminology. In Sec.~\ref{sec:density} we discuss a specific way of resolving time travel paradoxes due to Deutsch \cite{deutsch91}, highlighting that this method interprets mixed quantum states as ontic states. We then demonstrate that applying consistency conditions to epistemic states as if they were ontic states has the tendency to conceal paradoxes, rather than resolve them. In Sec.~\ref{sec:epistemic}, we outline some considerations that an epistemic interpretation of quantum mechanics must address and then examine the effect of CTCs on a toy epistemic model.

\section{Paradoxes with closed timelike curves}
\label{sec:paradoxes}

One of the challenges when discussing time travel is nomenclature, so we begin by introducing some terminology. A \textit{chronology-respecting} trajectory is a timelike worldline that never intersects itself. A \textit{chronology-violating} trajectory is a timelike worldline that intersects itself. A chronology-respecting (-violating) object is an object that is on a chronology-respecting (-violating) trajectory. According to a chronology-respecting observer, chronology-violating and chronology-respecting trajectories come together in some region in space-time where the systems on those trajectories can interact. Then the trajectories diverge and there are no more interactions between those systems. We consider networks for which there is only one chronology-violating worldline, namely, a CTC. 

Chronology violations make the chronological ordering of two time-like separated events, A and B, subjective.  This makes the notions of `past' and `future' ambiguous. (More precisely, within a special relativistic setting, it confuses the notions of past and future light cones.) To avoid this technical difficulty, all references to time are with respect to a chronology-respecting observer.

\begin{figure}
    \centering
    \includegraphics[width=0.4\textwidth]{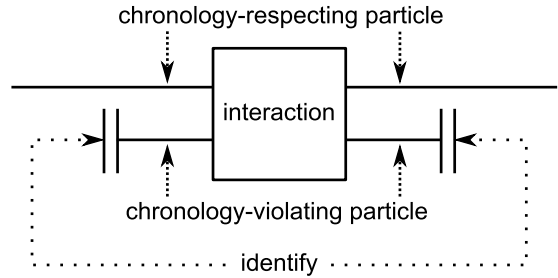}
    \caption{A chronology-violating network with an interaction between a chronology-respecting system and a chronology-violating system. In diagrams of this form, time flows from left to right. The double vertical lines are identified as a single spacetime point, forming a CTC.}
  \label{fig:circuitmodel}
\end{figure}

To examine the possible consequences of CTCs, we use the phenomenological circuit model introduced by Deutsch and illustrated in Fig.~\ref{fig:circuitmodel}. In diagrams of this type, time flows from left to right according to some chronology-respecting observer. The existence of a CTC is introduced by identifying two points in the diagram (denoted by double vertical lines) to form a closed loop, while the mechanism that generates the CTC is ignored. Each horizontal line in the diagram represents the worldline of a single particle, where worldlines terminating in double vertical lines represent chronology-violating particles, while worldlines extending to past- and future- infinity represent chronology-respecting particles. All interactions between chronology-respecting and chronology-violating particles, as well as the free evolution of all particles, are localized into a single equivalent interaction, represented by a box, acting on the ontic states of the particles.

From the perspective of classical physics, time travel can lead to paradoxes. We distinguish two classes of paradoxes: \textit{consistency paradoxes} and \textit{information paradoxes}. Due to the difficulty of defining `information' without reference to specific ontic and epistemic states, we will not discuss information paradoxes\footnote{An example of an information paradox is a scenario in which a physicist is given the plans to build a time machine and then uses those plans to go back in time and give himself the plans. In this case, the information required to build a time machine has spontaneously come into existence on a CTC.}. We define a \textit{consistency paradox} as a scenario for which there exist initial states of the chronology-respecting system with no consistent solution for the state of the chronology-violating systems. An example of a consistency paradox is the autoinfanticide paradox, in which an agent travels back in time and kills a younger version of themself before that version can travel back in time. This paradox implies that any agent who will choose to kill a younger version of themself and who possesses the ability to kill a younger version of themself cannot travel back in time. That is, the initial conditions are restricted because otherwise the causal loop would be inconsistent.

\section{Deutsch's consistency condition}
\label{sec:density}

Deutsch considered paradoxical situations, like the consistency paradox of the previous section, using quantum systems instead of classical ones~\cite{deutsch91}.  We now review his construction.

\begin{figure}
    \centering
    \includegraphics[width=0.35\textwidth]{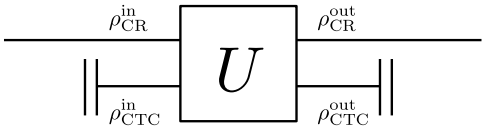}
    \caption{A chronology-violating network with an interaction $U$ between a chronology-respecting system and a chronology-violating system. The chronology-respecting (-violating) system is in the ontic state $\rho^{\rm in}_{\rm CR}$ ($\rho^{\rm in}_{\rm CTC}$) before the interaction and $\rho^{\rm out}_{\rm CR}$ ($\rho^{\rm out}_{\rm CTC}$) afterwards.}
    \label{fig:densitycircuit}
\end{figure}

Consider the circuit depicted in Fig.~\ref{fig:densitycircuit}, in which a chronology-violating system interacts with a chronology-respecting system. Because the chronology-respecting system has not interacted with any systems traversing the CTC in its unambiguous past (as defined by its own worldline), then the chronology-respecting and chronology-violating systems are unentangled immediately prior to the interaction (since they have had no common past in which to become entangled), i.e.,
\begin{equation}
    \rho^{\rm in}_{\rm CR} \otimes \rho^{\rm in}_{\rm CTC}\,.
\end{equation}
(Note that while the chronology-respecting and chronology-violating systems are assumed to be unentangled, the chronology-respecting system may be entangled with some other systems, so $\hat{\rho}^{\rm in}_{\rm CR}$ may represent a reduced density matrix of a larger system.) Therefore, after the two systems interact, the joint state of the system is
\begin{equation}
    U\left(\rho^{\rm in}_{\rm CR} \otimes \rho^{\rm in}_{\rm CTC}\right)U^{\dagger}\,,  \label{eq:onticout}
\end{equation}
where $U$ is an arbitrary unitary matrix representing the interaction between the two systems. In general, Eq.~(\ref{eq:onticout}) cannot be represented as a product state, i.e.
\begin{equation}
     U\left(\rho^{\rm in}_{\rm CR} \otimes \rho^{\rm in}_{\rm CTC}\right)U^{\dagger} \neq \rho_{1} \otimes\rho_2\,,
\end{equation}
for any two density matrices $\rho_1$ and $\rho_2$.

After the interaction, the two systems are separated, with the chronology-respecting system following its timeline into the unambiguous future, and the chronology-violating system traveling back in time. If the two systems remain entangled, then from the ontic interpretation of the quantum state advocated by Deutsch, the chronology-violating system can be affected by performing measurements on the chronology-respecting system. This would then update the state of the chronology-respecting system, which changes the way that the chronology-violating system is changed, and so on \textit{ad infinitum}. At this point, at least two possible solutions to this problem exist. The first is to use the notion of time-displaced entanglement~\cite{ralph}, treating entanglement between particles at different times as physically meaningful.

The alternative approach advocated by Deutsch corresponds to assuming that the means by which the systems are separated and the second system continues along the CTC (e.g., by one entering the mouth of a wormhole) causes an instantaneous change on both systems such that the output state takes the unentangled form $\rho^{\rm out}_{\rm CR} \otimes \rho^{\rm out}_{\rm CTC}$, where
\begin{align}
    \rho^{\rm out}_{\rm CR} &= {\rm Tr}_{\rm CTC} [U\left(\rho^{\rm in}_{\rm CR} \otimes \rho^{\rm in}_{\rm CTC}\right)U^{\dagger}]    \,, \label{eq:densitytrace} \\
    \rho^{\rm out}_{\rm CTC} &= {\rm Tr}_{\rm CR} [U\left(\rho^{\rm in}_{\rm CR} \otimes \rho^{\rm in}_{\rm CTC} \right)U^{\dagger}]\,. \label{eq:densityontic}
\end{align}
Although the partial trace is commonly used to describe subsystems, it usually does not signify a change in the joint state of the combined system.  Here, the partial trace represents a nonlocal physical mechanism that ensures that any entanglement between the chronology-respecting and violating qubits is removed.  Without analysing such a mechanism in detail, we note the similarities to some quantum descriptions of the process of black hole evaporation.  

Finally, Deutsch identifies the state of the chronology-violating systems after the interaction with the state of these systems before, requiring that
\begin{equation}
  \label{eq:densityonticconstraint}
  \rho^{\rm in}_{\rm CTC} = \rho^{\rm out}_{\rm CTC} \,.
\end{equation}
Deutsch denotes this as the \emph{kinematical consistency condition}.  It arises because, with this interpretation, $\rho_{\rm CTC}$ is the element of reality that travels back in time.  The CTC is identifying the physical state of affairs at two points, ``out'' and ``in'', and thus if $\rho_{\rm CTC}$ is an ontic state then it is a sensible requirement to enforce the kinematical consistency condition (\ref{eq:densityonticconstraint}).

With this condition, Deutsch showed that there are never any time travel paradoxes, because Eq.~(\ref{eq:densityontic}) subject to the constraint of Eq.~(\ref{eq:densityonticconstraint}) always possesses a solution. (In general, it possesses many solutions, and Deutsch proposes a maximum-entropy condition to select a unique one that corresponds to the physical state of the system.) However, it arises from a non-standard interpretation of the generally mixed state $\rho_{\rm CTC}$ as an ontic state, whereas mixed states are typically interpreted as epistemic states. 

\begin{figure}
  \centering
    \includegraphics[width=0.4\textwidth]{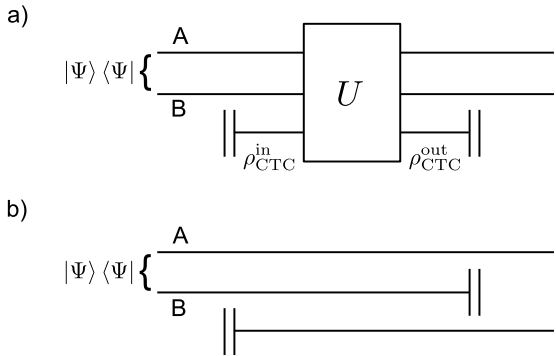}
  \caption{a) One half of a chronology-respecting Bell pair interacting with a qubit on a CTC. b) One half of a Bell pair entering a time machine at a time $t$ to travel back to a time $t-\delta t$ (according to some chronology-respecting observer).}
  \label{fig:swapquantum}
\end{figure}

To see that the partial trace condition (\ref{eq:densityonticconstraint}) corresponds to a nonlocal physical change that removes correlations between chronology-respecting and chronology-violating systems, consider the situation depicted in Fig.~\ref{fig:swapquantum}~(a) for quantum particles using Deutsch's consistency condition. Systems $A$ and $B$ are qubits prepared in some maximally entangled state $\hat{\eta}_{AB}$ (where the subscripts denote the systems that are entangled) and the chronology-violating qubit is in some mixed state $\hat{\rho}_C$. The joint state of the three qubits immediately before the interaction is
\begin{equation}
  \alpha = \hat{\eta}_{AB}\otimes\hat{\rho}_{C}	\,.
\end{equation}
The swap gate simply swaps the states of qubit $B$ and the chronology-violating qubit, so the joint state of the three qubits immediately after the interaction is
\begin{equation}
  \alpha = \hat{\eta}_{AC}\otimes\hat{\rho}_{B}	\,.
\end{equation}
Because the trace of any maximally entangled state over any subsystem is the maximally mixed state, the only consistent state of the chronology-violating qubit is 
\begin{align}
\rho_{C} &= \frac{1}{2}\mathbb{I}_2	\,,
\end{align}
i.e., the maximally mixed state.  (Here, $\mathbb{I}_2$ is the $2\times 2$ identity matrix.) From Eq.~\ref{eq:densitytrace}, the corresponding output of the chronology-respecting qubits is also the maximally mixed state. So not only is there no entanglement between chronology-respecting and chronology-violating qubits, but also all correlations between the two chronology-respecting qubits have been removed by this interaction under Deutsch's consistency condition and all information has been lost.

There are two equivalent ways of considering the interaction implemented by a swap gate. The first, depicted in Fig.~\ref{fig:swapquantum}~(a), corresponds to the internal states of two distinct particles being swapped, while the second, shown in Fig.~\ref{fig:swapquantum}~(b), corresponds to a physical swap of the particles, in which qubit $B$ enters a time machine at a time $t$ (with respect to a chronology-respecting observer), emerging at some earlier time $t-\delta t$ and continuing on towards future infinity. In this second picture, no interaction has actually taken place and, by assumption, no non-unitary evolution has occurred. In the limit $\delta t\rightarrow 0$, nothing actually happens physically, and yet all correlations and information are lost.\footnote{An explanation of this phenomenon within the Everett interpretation is that that time travel always leads to particles traveling between branches of the `universal wave-function' and so particles within a particular branch are no longer correlated because they were not prepared in an entangled state together (i.e., they were prepared in entangled states with copies of each other). However, this explanation obviously only applies to the Everett interpretation, as it requires branches of some `universal wave-function' for particles to travel between.}

\subsection{Revisiting Deutsch's consistency condition}
\label{sec:classicalepistemic}

Is the kinematical consistency condition a reasonable consistency condition to apply to quantum states? If a particle travels back in time, it is natural to equate the physical properties of the particle at the point ``out'' with those at the point ``in'' (given that we localized all evolution of the time-traveling particle to the interaction). With this in mind, the kinematical consistency condition (\ref{eq:densityonticconstraint}) relies on the mixed state $\rho_{\rm CTC}$ being an ontic state, i.e., an element of reality. This is an unconventional interpretation.

In this section, we argue that consistency conditions such as Eq.~(\ref{eq:densityonticconstraint}) cannot be immediately applied to epistemic states, rather, they must be applied to ontic states. This is a critical point, because it illustrates that unless mixed states are ontic states, the kinematical consistency condition is not an appropriate consistency condition and consequently any predicted behavior that results from applying it to quantum systems near CTCs can be interpreted as resulting from an observer failing to distinguish their knowledge about reality from reality itself. This feature was recognized by Deutsch, who stated that the kinematical consistency condition is (emphasis his)~\cite{deutsch91}
\begin{quotation}
the correct condition under the unmodified quantum formalism, \textit{but it is either wrong or insufficient under every other version of quantum theory,} just as under classical physics.
\end{quotation}
Note that, to Deutsch, the unmodified quantum formalism is the Everett interpretation. 

To demonstrate that consistency conditions must be applied to ontic states, we present a classical example where there are no consistent solutions for the ontic state (i.e., a classical time travel paradox), but an observer with limited knowledge would not see the paradox if they apply consistency conditions to their knowledge of the classical system, rather than to the state of the classical system itself.

Consider a classical object with 2 states (i.e., a bit) traveling on a CTC. It is initially (with respect to some chronology-respecting observer) in either the ontic state 0 or in the ontic state 1. A bit-flip operation, $F$, is then applied, so that if it was in the ontic state 0 it is now in the ontic state 1 and \textit{vice versa}. But then when the bit continues on the loop and arrives back at the same point, it is in the opposite ontic state. This is a consistency paradox for either of the possible initial ontic states, since the physical state of a system should be unique (single-valued) at any point in spacetime.

Now consider the epistemic state that an observer assigns to the bit. For a classical bit, the epistemic state takes the form of a probability distribution over the possible ontic states. If the observer is uncertain about the ontic state of the bit and describes the bit immediately prior to the flip operation by the epistemic state
\begin{align}
    P^{\rm in}(0) &= 0.5  \,,\\ 
    P^{\rm in}(1) &= 0.5 \,,
\end{align}
then after the bit flip, the observer would assign the epistemic state
\begin{align}
    P^{\rm out}(0) &= P^{\rm in}(F(1)) = 0.5    \,,\\
    P^{\rm out}(1) &= P^{\rm in}(F(0)) = 0.5   \,,
\end{align}
to the system, which is the same epistemic state as before. The observer's knowledge is consistent and consequently the observer would describe the situation as nonparadoxical even though there is no valid ontic state assignment for the bit. The self-consistency only occurs because the observer is implicitly treating an epistemic state as an ontic state when assessing the consistency of the epistemic state. That is, applying consistency conditions to epistemic states \emph{hides} paradoxes from the observer, rather than resolving them. In this particular case, an observer having a consistent description of the system does not guarantee that there is a consistent ontic state that the system can be in.

In this example, the classical ontology makes it easy to see that requiring the epistemic states before and after the interaction to be the same is an inappropriate consistency condition; instead, a consistent ontic state assignment should be required. In the quantum case, there is an ambiguity due to the lack of a clear, unambiguous ontology of the theory. However, by analogy to the classical case, if all quantum states are epistemic states and there is some as yet unknown underlying ontology, then the kinematical consistency condition will not be appropriate.

\section{Epistemic states and Closed Timelike Curves}
\label{sec:epistemic}

We have shown that imposing certain consistency conditions on epistemic states (implicitly treating them as ontic states) can conceal paradoxes by allowing consistent \textit{epistemic} state assignments to situations for which there is no consistent \textit{ontic} state assignment. We take the perspective that all quantum states, both pure and mixed, are epistemic. Consequently, we view the kinematical consistency condition as inappropriate. We would therefore like to consider how an epistemic theory of quantum states would function in the presence of CTCs. Before we examine this, we review the properties of an epistemic theory.

\subsection{Epistemic interpretations of quantum mechanics}

There are two main approaches to viewing quantum states as epistemic, which differ primarily in their interpretation of what the knowledge represented by a quantum state pertains to. One is the so-called "radical Bayesian" approach~\cite{fuchs03}, in which quantum states catalog the personal betting odds that a rational agent assigns to the outcomes of future measurements. Within this approach, there is no reference to an underlying ontology, and the only clear consistency conditions would be to enforce Dutch-book consistency (i.e., consistency of betting odds). We do not pursue this approach further (in part due to the problems of assigning betting odds with time-traveling parties), but note that it has been argued~\cite{cavalcanti} that Deutsch's nonlinear evolution is inconsistent by this criterion.

The second approach to an epistemic perspective of quantum states is to consider them as describing an observer's knowledge about some underlying ontology (often called ``hidden variables''). That is, quantum states correspond to an observer's knowledge about an objective state of the universe. Any such ontological model that reproduces the predictions of quantum mechanics is necessarily \emph{non-local}. In addition, one could demand (as we do) that the wavefunction in an epistemic theory must correspond \emph{only} to an observer's knowledge and not to any reality. Such an epistemic theory is said to be $\Psi$-epistemic \cite{harrigan07}. (In contrast, Bohm-deBroglie theory provides a non-local ontological model that reproduces quantum mechanics, but it is a $\Psi$-ontic theory wherein the wavefunction takes the form of a physical potential, and so is not fully compatible with the epistemic perspective that we advocate.) 

Without a satisfactory ontological model for a $\Psi$-epistemic interpretation of quantum mechanics, it is difficult to determine consistency conditions that are appropriate in the presence of CTCs. To gain some understanding of this issue, we consider how Spekkens' toy epistemic theory \cite{toy} functions in the presence of CTCs. While the toy theory is a deliberately incomplete model of quantum mechanics, it reproduces many so-called `quantum' phenomena, such as incomplete information from any measurement, the necessity of disturbance of the system associated with measurement, mutually unbiased bases, remote steering, no-cloning theorems, quantum teleportation and superdense coding. In addition, it has a clearly defined ontology. The toy theory therefore provides a reasonable starting point to look at how quantum states, considered purely as states of knowledge, would behave in the presence of CTCs. We now briefly review some properties of the toy theory; for full details, see~\cite{toy}.

\subsection{Epistemic and ontic states within a toy theory}

Proponents of an epistemic perspective of quantum mechanics view quantum states as describing an observer's knowledge. The fundamental question that an epistemic interpretation must answer is, what does an observer have knowledge about?

In Spekkens' toy theory~\cite{toy}, the answer to this question is that it is knowledge about a hidden classical ontology and there is a fundamental constraint on the amount of knowledge the observer may possess. The constraint on the observer's knowledge is the \textit{knowledge-balance principle}, which states that an observer can never have more than half of the information required to completely specify the current ontic state of the system. That is, an observer's knowledge about the ontic state of a system is restricted. For the purposes of developing a simple model, an observer's \textit{knowledge} about a system is defined to be the number of `yes-no' questions they can answer about the ontic state of the system from a canonical set of questions, where a canonical set of questions contains only the smallest number of questions needed to specify an arbitrary ontic state. The \textit{epistemic state} that an observer assigns to a system is the set of ontic states of the system that are consistent with their knowledge.

This formulation of knowledge as answers to binary questions produces epistemic states that are analogous to qubits. To illustrate the definitions of `knowledge', `epistemic states' and `canonical sets of questions', consider a set of 4 ontic states,
\begin{equation}
\{1,\ 2,\ 3,\ 4\}   \,.
\end{equation}
One (of the many possible) canonical set of yes-no questions is `is the system in one of the ontic states 1 or 2 (denoted by $1\vee2$) or not?' and `is the system in one of the ontic states $1\vee4$ or not?' The corresponding epistemic states are $1\vee2$ and $3\vee4$ (depending on the answer to the first question), and $1\vee4$ and $2\vee3$ (depending on the answer to the second question). For the knowledge balance principle to be obeyed for the above set of ontic states, an arbitrary observer can answer at most one of the two questions needed to completely specify the ontic state. In the above example, an observer who can answer at most one of the given questions can describe the ontic state as $1\vee2$, $1\vee4$, $3\vee4$, $2\vee3$ or $1\vee2\vee3\vee4$. Note that the question `is the system in the ontic state 1 or not?' is not a canonical question, as it does not provide a maximally efficient means of partitioning the space of ontic states.

\begin{figure}
    \centering
        \includegraphics[width=0.4\textwidth]{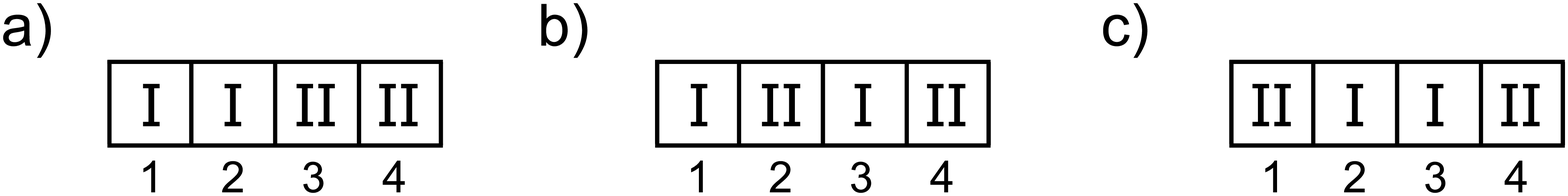}
    \caption{Partitioning of the ontic space corresponding to measurements that distinguish between the epistemic states analogous to the quantum states: a) $|0\rangle$ and $|1\rangle$, b) $|+\rangle$ and $|-\rangle$ and c) $|i\rangle$ and $|{-}i\rangle$, where I denotes an ontic state that is in the ontic support of the first epistemic state in the pair and II represents one in the ontic support of the second.}
    \label{fig:partition}
\end{figure}

The set $\{1,2,3,4\}$ is the smallest set of ontic states that allow observers to possess some knowledge (that is, for an observer to answer at least one question about the ontic state of the system) while not violating the knowledge balance principle. In addition, epistemic states occur in orthogonal pairs, so this set is identified as a \textit{toybit} (analogous to a qubit). The three ways that an observer can partition the ontic space of a toybit by asking one question (i.e.,~performing a `measurement') are shown in Fig.~\ref{fig:partition}. The \textit{ontic support} of an epistemic state is the set of all ontic states consistent with that epistemic state. For example, the ontic support of the epistemic state $1\vee2$ is the set $\{1,2\}$. Two epistemic states are orthogonal in the toy theory if the intersection of their ontic supports is the empty set, while two epistemic states are \textit{compatible} (i.e. there are physical situations that can be consistently described by both epistemic states) if the intersection of the ontic supports is non-empty. For example, the epistemic states $1\vee2$ and $3\vee4$ are orthogonal while the epistemic states $1\vee2$ and $1\vee4$ are compatible.

If `measurements' do not disturb the ontic state of the system, then an observer could violate the knowledge balance principle. In the above example, performing a measurement that reports if the toybit is in one of the ontic states $1\vee2$ or not, and then performing a second measurement to determine whether the toybit is in one of the ontic states $1\vee3$ or not would imply that the ontic state of the system would be completely known (e.g., if both outcomes are `yes', then the ontic state must be $1$). To maintain the knowledge-balance principle, a measurement update rule is introduced, so that when a measurement is performed upon a system and the system is observed to be in the epistemic state with ontic support $\{n_1,n_2,...,n_j\}$, then the ontic state of the system is randomly disturbed to a post-measurement ontic state sampled uniformly from this support $\{n_1,n_2,...,n_j\}$.  Therefore `measurements' in the theory are statistical in a similar manner to measurements in quantum mechanics.

\begin{figure}
    \centering
        \includegraphics[width=0.4\textwidth]{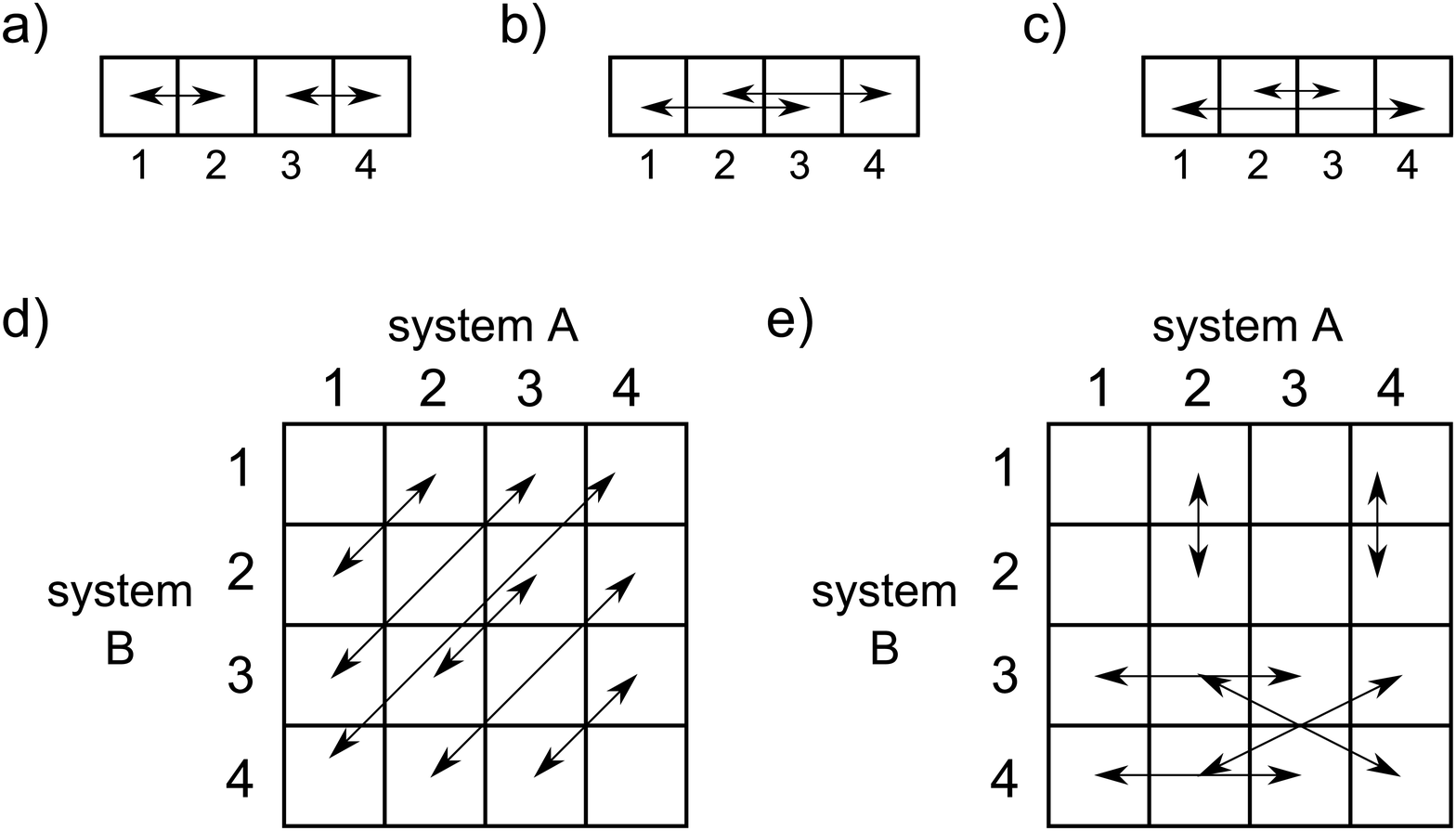}
    \caption{Representations of transformations on a single toybit corresponding to the following elements of $S_4$ in standard cycle notation: a) (12)(34), b) (13)(24) and c) $(14)(23)$. These transformations are analogous to the qubit transformations $\sigma_z$, $\sigma_x$ and $\sigma_y$ respectively, since if an observer applies the transformation (a) to a system which they describe by one of the epistemic states in Fig.\,\ref{fig:partition}~(a), the epistemic state will still be a valid description. Representations of two system interactions in the toy theory, analogous to the: d) SWAP and e) C-NOT gates in quantum mechanics. The transformation in (e), together with single system transformations, generates the group of valid two system interactions in the toy theory.}
    \label{fig:toytransformations}
\end{figure}

The last feature of the toy theory needed to study CTCs are valid transformations. A valid, reversible transformation of a system within the toy theory is a permutation of ontic states; such a transformation induces a permutation of epistemic states.  For example, (12)(34) denotes the permutation that acts on a single toybit, swapping ontic states $1$ and $2$ and simultaneously swapping the ontic states $3$ and $4$; (1234) denotes a 4-cycle. For a single toybit, the set of valid transformations is S$_4$, the group of permutations of four objects. Some transformations on a toybit, together with all transformations on two toybits that are used in this work are shown diagrammatically in Fig.\,\ref{fig:toytransformations}.

\subsection{Consistency conditions on toybits traversing closed timelike curves}

We now consider how toybits behave in the presence of CTCs, because with a clear ontology it is straightforward to apply consistency conditions on the ontic states. We begin by presenting our ontic consistency condition in the toy theory and demonstrating that paradoxes can occur. We also demonstrate that CTCs introduce initial and final boundary conditions in the toy theory.

\begin{figure}
  \centering
  \includegraphics[width=0.4\textwidth]{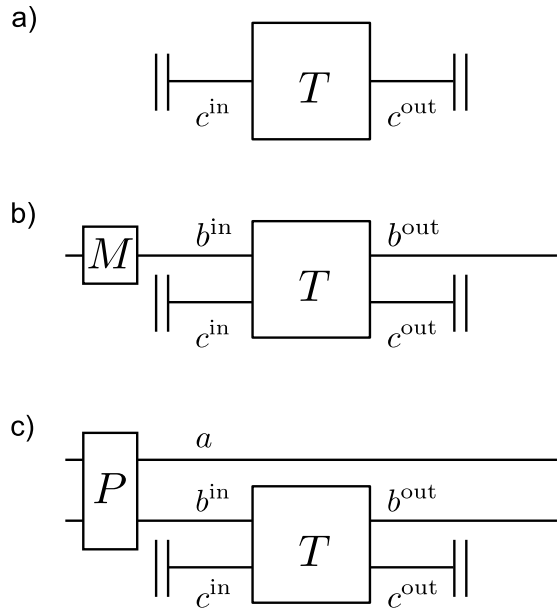}
  \caption{a) Circuit of a toybit on a CTC being acted upon by a transformation $T_{1}$. b) Circuit of a chronology-respecting toybit that has a measurement $M$ performed on it and then interacts with a chronology-violating toybit. c) Circuit of a chronology-respecting toybit being prepared in a correlated epistemic state with another chronology-respecting toybit $a$ via a state preparation procedure $P$ and then interacting with a system on a CTC through the interaction $T_{2}$. The letters on the worldlines denote the ontic state of the corresponding system, with the superscript indicating whether the system is in the input or output state.}
  \label{fig:toycircuit}
\end{figure}

The natural consistency condition is that the ontic state of the chronology-violating toybit is identified at the points ``out'' and ``in''. For the toy theory, this equates to
\begin{equation}
    c^{\rm out} = c^{\rm in}    \,,  \label{eq:ontcon}
\end{equation}
where $c^{\rm out}$ ($c^{\rm in}$) is the ontic state of the chronology-violating toybit depicted in Fig.\,\ref{fig:toycircuit}~(a) after (before) the interaction.

Applying this condition to the circuit in Fig.\,\ref{fig:toycircuit}~(a), where the transformation $T_1$ maps ontic states according to the permutation (1234), one can immediately see that there are no ontic states that satisfy Eq.~(\ref{eq:ontcon}) and so this transformation would lead to a time travel paradox. We note that any transformation that does not possess a fixed point must necessarily be paradoxical. The existence of paradoxes here is analogous to the classical situation, in which paradoxes can also occur, and can be understood as arising from the underlying classical ontology in this theory. It is, however, in direct contrast to the result obtained in the quantum case using the kinematical consistency condition proposed by Deutsch, for which no paradoxes could occur. 

Now consider the transformation (123)(4). In this case, there is a fixed point of the transformation, namely the ontic state 4, so this transformation does not give rise to a paradox. However, the ontic state 4 is the only fixed point of the transformation. Therefore an observer could say that the chronology-violating system is definitely in the ontic state 4, violating the knowledge balance principle. All of the quantum mechanical phenomena that are reproduced in the toy theory are reproduced because of the restriction on an observer's knowledge. Therefore, the reproduction of generic quantum behavior may break down when toybits traverse a CTC. This is reminiscent of the breakdown of some of the rules of quantum mechanics (e.g. linear evolution of states, global unitarity) that occur when using Deutsch's consistency condition.

In principle this may not be an issue provided that violations of the knowledge balance principle can only occur sufficiently close to a CTC (i.e. if there is a closed horizon outside of which violations of the knowledge balance principle cannot occur), because from an epistemic perspective there is no \textit{a priori} reason to require that the evolution of particles along CTCs is the same as the evolution of chronology-respecting particles. However, if an observer can use a CTC to violate the knowledge balance principle for chronology-respecting toybits, then the behavior of the toy theory would be qualitatively different depending on whether or not observers have access to CTCs.

To examine the knowledge balance principle for chronology-respecting toybits in the presence of CTCs, we consider a chronology-respecting toybit interacting with a chronology-violating toybit as shown in Fig.~\ref{fig:toycircuit}~(b). Due to the difficulty in characterizing transformations of three or more toybits, we will only consider a single chronology-respecting toybit interacting with a single chronology-violating toybit. 

If the transformation $T_2$ in Fig.~\ref{fig:toycircuit}~(b) is a product of local transformations, then the interaction may or may not be paradoxical (depending on whether the transformation of the chronology-violating toybit has fixed points or not), but it will not lead to a violation of the knowledge balance principle. The group of two toybit transformations is generated by local transformations and the correlating transformation $T_{CN}$ in Fig.~\ref{fig:toytransformations}~(e), which is analogous to a CNOT gate. Composing $T_{CN}$ with local transformations will not change whether or not the knowledge balance principle is violated, so we need only consider the case $T_2=T_{CN}$. 

For this case, there are ontic states of the joint system that satisfy Eq.~(\ref{eq:ontcon}), namely, whenever the ontic state of the chronology-respecting toybit is $1$ or $3$ (where we have taken the chronology-respecting system to be system A). If, however, the observer measures the chronology-respecting toybit before the interaction and finds, for example, that it is in one of the states $1\vee2$, then the only ontic state of the chronology-respecting toybit that allows for a consistent solution of Eq.~(\ref{eq:ontcon}) is the state $1$. The observer has therefore obtained complete knowledge of the chronology-respecting toybit that is correct from the time the measurement is made until the transformation is applied. After the transformation is applied, the chronology-respecting system can be in either of the ontic states $1$ (if the ontic state of the chronology-violating toybit is $1$ or $2$) or $3$ (if the ontic state of the chronology-violating toybit is $3$ or $4$), and so the knowledge balance principle is obeyed after the interaction.

This process may appear to violate the knowledge balance principle in the intermediate time between the measurement and the transformation. However, even though the observer can completely specify the ontic state of the system at a given time, this does not constitute a violation of the knowledge balance principle as the observer's knowledge is contingent upon a future interaction. 

Similar behavior can occur without involving CTCs in scenarios involving pre- and post-selection. For example, consider a toybit prepared in an initial (epistemic) state $1\vee 2$. An observer can define conditional post-selection probabilities $p(o|M=m,E)$ that the current ontic state is $o$ given that the outcome $m$ will be observed in a future measurement $M$ on a system prepared in the epistemic state $E$. If the future measurement is $1\vee 3$ versus $2\vee 4$, then, contingent on the future outcome being $1\vee3$, the observer knows, with ``certainty'' that the current ontic state is $1$ (equivalently, if they perform the measurement and observe the outcome, they can retrodict with certainty the ontic state of the toybit before the measurement is performed). However, as discussed in Ref.~\cite{toy,liouville}, this is not a contradiction of the knowledge balance principle, correctly formulated, as the knowledge balance principle only applies to an observer's present knowledge and the observer does not know which outcome will occur. In this example, the observer's epistemic state, contingent upon the measurement being performed but \textit{without} post-selecting on the outcome is still $1\vee2$ as the probability that they assign to the system being in the ontic state $o$ is
\begin{align}
p(o|E) = \sum_m p(o|M=m,E)p(M=m|E)	\,,
\end{align}
where $p(M=m|E)$ is the probability that the observer assigns to the outcome $m$ of the measurement $M$ when performed on a system prepared in the epistemic state $E$. This scenario is directly analogous to the `violation' of the uncertainty principle in pre- and post-selection scenarios in quantum mechanics~\cite{liouville}.

The scenario in Fig.~\ref{fig:toycircuit}~(b) is equivalent to post-selecting on an ontic state that allows for nonparadoxical evolution for the specified transformation. However, if instead of $T_{CN}$, the interaction was the transformation $T'_2$, obtained by composing the permutation $(12)$ acting on the chronology-respecting toybit with $T_{CN}$, then the ontic state of the chronology-respecting system between the time of measurement and the interaction with the chronology-violating toybit must be $2$. If the observer cannot, even in principle, know which of the two interactions described above will occur, then from the time of measurement to the interaction, the correct epistemic state they should assign is $1\vee 2$, as they should set the probability $p(o)$ of the system being in the ontic state $o$ to
\begin{align}
p(o) = p(T_2) p(o|T_2) + p(T'_2) p(o|T'_2)	\,,
\end{align}
where $p(T'_2) = p(T'_2) = 1/2$ and $p(o|T)$ is the probability of the ontic state being $o$ conditioned upon the future interaction being $T$. Therefore the knowledge-balance principle can be maintained for chronology-respecting systems by restricting an observer's knowledge about how chronology-violating systems interact with chronology-respecting systems.

Note that even if the observer does not know how a chronology-violating system will interact with a chronology-respecting toybit, the interaction still places a final boundary condition on toybits. If the interaction could somehow be changed (e.g., if the permutation $(12)$ can be applied to the chronology-respecting toybit), then this has a retrocausal effect, in that the ontic state of the system between the measurement and the transformation $(12)$ must be changed to avoid a paradox.

In addition to placing final boundary conditions on toybits, CTCs can also introduce initial boundary conditions of toybits that interact with chronology-violating particles. For example, if the chronology-respecting and chronology-violating particles represented by Fig.~\ref{fig:toycircuit}~(b) interact via the transformation in Fig.~\ref{fig:toytransformations}~(e), then the only way Eq.~(\ref{eq:ontcon}) can be satisfied is if the chronology-respecting system is initially in either the ontic state 1 or the ontic state 3. This behavior matches the predictions of classical physics, where initial and/or final boundary conditions are often required to avoid paradoxes. 

\subsection{Consistency conditions for epistemic states}

We have shown that the ontic states of the toy theory in the presence of CTCs can be paradoxical, just like the classical case and unlike the quantum behavior that follows from Deutsch's kinematical consistency condition. However, as we will show, the epistemic states of the toy theory in these scenarios can serve to conceal the paradoxes in the underlying ontological model. This provides us with a new perspective on Deutsch's result.

In the toy theory, an observer's epistemic state is a probability distribution over the ontic states that is uniform and non-zero for the ontic states that are consistent with the observer's knowledge and zero for those ontic states that are inconsistent with the observer's knowledge. In the presence of CTCs, an observer's additional knowledge about the dynamics of a system on a CTC only allows them to say what ontic states are consistent. Of the states that are consistent, the observer has no reason to prefer one state over another (since there is no preferred state in the toy theory), and so should assign equal probabilities to the ontic states that are consistent with the interaction and zero to the states that are not consistent with the interaction.

Consider the scenario depicted in Fig.~\ref{fig:toycircuit}~(a) (analogous to the one in Sec.\,~\ref{sec:classicalepistemic}), in which a toybit on a CTC has the transformation (13)(24) applied to it. There is no ontic state that the toybit can consistently be in and so the situation should be seen as paradoxical. However, there are three epistemic states ($1\vee3$, $2\vee4$ and $1\vee2\vee3\vee4$) that are invariant under the transformation $(13)(24)$, and so if an observer enforced a consistency condition upon their knowledge, there would be a consistent solution to the evolution of epistemic states, even though there is no consistent ontic state assignment. That is, applying a direct analog of Deutsch's consistency condition to the epistemic states would \emph{conceal}, rather than \emph{resolve}, paradoxical ontic state assignments.

Moreover, consider the zero knowledge epistemic state, $1\vee2\vee3\vee4$. This epistemic state is invariant under any valid transformation on a single toybit, since all transformations in the toy theory are permutations of ontic states. Therefore it would satisfy an analog of Deutsch's consistency condition for any transformation acting on a single chronology-violating toybit. However, this epistemic state does not resolve any paradoxes for the ontic states. Rather, it corresponds to an observer knowing nothing and consequently not knowing that anything is wrong. 

From an epistemic interpretation, then, Deutsch's proof that there is always a consistent, but not necessarily pure, density state assignment to qubits on CTCs is no surprise. With an epistemic perspective it often corresponds to an observer not having enough information to discern that there is in fact a paradox. For our example of a CNOT coupling, if one applied a condition analogous to Deutsch's maximum entropy condition, namely, that observers should assign the state of minimum knowledge (equivalent to maximum entropy from an epistemic perspective) that is self-consistent, then observers will never be able to say anything about the dynamical behavior of toybits traversing CTCs, as the zero-knowledge state is always consistent. If there is some underlying ontology in quantum mechanics, then such concealment of paradoxes may occur under the consistency conditions that have been considered.

\subsection{Correlations between chronology-respecting and violating systems}

One of the issues that arises when quantum mechanics is applied to particles traversing CTCs is the possibility of time-displaced entanglement. Under an ontic interpretation of any quantum states (either wavefunctions or density matrices), time-displaced entanglement would allow a measurement on a chronology-respecting system to change the state of a qubit on a CTC that is in the causal history of the chronology-respecting system. As shown in Sec.~\ref{sec:density}, Deutsch's consistency condition not only excludes the possibility of entanglement between the CTC qubit and chronology-respecting qubits by using the partial trace, i.e., by physically changing the ontic state of the system, but also acts to reduce entanglement between chronology-respecting systems that interact with chronology-violating systems. 

We now explore how an analogous situation is handled in the toy theory, which is defined as a strictly local theory. One advantage of an epistemic perspective is that time-displaced entanglement can be viewed as a form of (noncausal and nonclassical) correlation. In the strictly local toy theory, when an observer measures one half of a pair of systems that they describe by some correlated epistemic state, then the outcome of the measurement allows them to update the epistemic state by which they describe the other system, without any physical change occurring in the other system. These correlations are sufficient to allow for `quantum' teleportation, superdense coding and remote steering, and so they seem to capture at least some of the quantum correlations. While, from an epistemic perspective, other correlations (especially those required to violate a Bell inequality) must be due to a nonlocal ontology, examining how the local correlations behave in the presence of CTCs will reveal something of how correlations between epistemic quantum states may behave.

Consider the circuit shown in Fig.~\ref{fig:toycircuit} (c), where the transformation is analogous to a swap gate, which acts on ontic states as shown in Fig.~\ref{fig:toytransformations} (d). Systems A and B are chronology-respecting toybits (in the ontic states $a$ and $b$ respectively), described by some observer by the maximally correlated epistemic state 1.1 v 2.2 v 3.3 v 4.4 before the interaction. System C (in the ontic state $c$) is a chronology-violating toybit that interacts with toybit B via the swap gate. Since toybits B and C interact via the swap gate, the only consistent ontic solution is that toybit C must always be in the same ontic state that toybit B was in before the interaction and the state of toybits A and B are not changed by the interaction. Therefore, after the interaction, toybits A and B can be described by the same maximally correlated epistemic state. That is, the CTC does not decorrelate the systems.

While from this example everything looks sensible from an epistemic perspective, it is important to remember that the toy theory is a strictly local theory and it is unclear how any nonlocal variables (responsible for correlations that allow violations of Bell inequalities) would behave in the presence of CTCs. The swap gate in the toy theory acts only on the ontic state of toybit B, whereas in an epistemic interpretation of quantum theory, the swap gate would act on a nonlocal ontological state.

\section{Conclusion}
\label{sec:conclusion}

The consistency conditions appropriate to many interpretations of quantum mechanics will lead to paradoxical situations, just as in classical physics.  Deutsch argues for a particular interpretation, with mixed states being ontic, in order to avoid such paradoxes.  If, however, mixed quantum states are indeed epistemic, then Deutsch's treatment may simply be \emph{concealing} paradoxes because there is always a consistent epistemic state assignment (e.g., a zero knowledge state) to scenarios where there is no truly consistent ontic state assignment.

More specifically, if a mixed density matrix is taken to represent an observer's knowledge, then Deutsch's consistency condition (or any other consistency condition imposed upon the density matrix) will be inappropriate. This suggests that many of the results that have been derived from Deutsch's consistency condition may be due to an incorrect consistency condition rather than genuine physical behavior near CTCs.  This is a general consequence of attempting to apply the quantum formalism to scenarios arising in general relativity, which can be a challenge without a clearly defined ontology. 

In order to illustrate possible effects of CTCs from an epistemic perspective with a clearly defined ontology, we examined CTCs in the context of Spekkens' incomplete toy theory, which is based upon the principle that there is a fundamental restriction on knowledge. This principle allows for many quantum mechanical phenomena to occur in a classical ontology. In accordance with classical physics and in contrast to results based on Deutsch's consistency condition, we find that introducing CTCs into the toy theory and allowing arbitrary interactions between chronology-respecting and chronology-violating systems places both initial and final boundary conditions on the ontic state of systems that interact with CTCs and also leads to time travel paradoxes.

The toy theory also provides a physically reasonable way of treating correlations between systems such that not all information is lost when one half of an entangled pair interacts with a chronology-violating system (which is what happens when Deutsch's consistency condition is applied).

Introducing CTCs into the toy theory for two toybits can be done without violating the knowledge balance principle by enforcing an additional constraint on observers' knowledge, namely, that they cannot know in advance how chronology-violating and chronology-respecting systems will interact. However, this highlights an apparent retrocausality in the toy theory, in that if the interaction is not fixed in advance, then the ontic state of a chronology-respecting system may be altered by the choice of interaction in order to avoid a paradox. It is currently unclear whether or not the knowledge balance principle can be violated in the presence of CTCs for larger numbers of toybits.

In our opinion, the issues surrounding time travel have not been satisfactorily resolved. Rather, there is much work to be done to determine whether or not CTCs result in physically unreasonable behavior, but this work must be done in the context of a clearly defined ontology, which will in turn depend upon the interpretation of quantum mechanics that is employed.

\begin{acknowledgments}
We acknowledge helpful discussions with Eric Cavalcanti and Nick Menicucci. SDB acknowledges the support of the ARC and the Perimeter Institute for Theoretical Physics.  
\end{acknowledgments}


\begin{thebibliography}{18}

\bibitem{godel49} K.~G\"odel, Rev. Mod. Phys. \textbf{{21}}, 447 (1949).

\bibitem{gott91} J.~R.~Gott, Phys. Rev. Lett. \textbf{{66}}, 1126 (1991).

\bibitem{morris88} M.~S.~Morris, K.~S.~Thorne and U.~Yurtsever, Phys. Rev. Lett. \textbf{61}, 1446 (1988).

\bibitem{ori07} A.~Ori, Phys. Rev. D \textbf{76}, 044002 (2007).

\bibitem{tipler74} F.~J.~Tipler, Phys. Rev. D \textbf{9}, 2203 (1974).
  
\bibitem{deutsch91} D.~Deutsch, Phys. Rev. D \textbf{44}, 3197 (1991).

\bibitem{everett} H.~Everett, Rev. Mod. Phys. \textbf{29}, 454 (1957).
  
\bibitem{bacon04} D.~Bacon, Phys. Rev. A \textbf{70}, 032309 (2004).

\bibitem{aaronson08} S.~Aaronson and J.~Watrous, Proc. R. Soc. A \textbf{465}, 631 (2009).

\bibitem{brun08} T.~Brun, J.~Harrington and M.~Wilde, Phys. Rev. Lett. \textbf{{102}}, 210402 (2009).

\bibitem{frey} R.~DeJonghe, K.~Frey and T.~Imbo, Phys. Rev. D \textbf{81}, 087501 (2010).

\bibitem{durham} I.~Durham, arXiv:0803.3287v3 [quant-ph].
  
\bibitem{dasilva2010} R.~D.~da Silva, E.~F.~Galvao and E.~Kashefi, Phys. Rev. A \textbf{83}, 012316 (2011).

\bibitem{lloyd2010} S.~Lloyd, \textit{et al.}, Phys. Rev. Lett. \textbf{106}, 040403 (2011).

\bibitem{svetlichny2009} G.~Svetlichny \textit{et al.}, arXiv:0902.4898v1 [quant-ph].

\bibitem{bennett} C.~H.~Bennett, D.~Leung, G.~Smith and J.~A.~Smolin, Phys. Rev. Lett. \textbf{103}, 170502 (2009).
  
\bibitem{ralph} T.~C.~Ralph, Proc. SPIE \textbf{6305}, 63050P (2006).

\bibitem{fuchs03} C.~A.~Fuchs, J. Mod. Opt. \textbf{50}, 987 (2003).

\bibitem{cavalcanti} E.~G.~Cavalcanti and N.~C.~Menicucci, arXiv:1004.1219v4 [quant-ph].
  
\bibitem{harrigan07} N.~Harrigan and R.~W.~Spekkens, Found. Phys. \textbf{40}, 125 (2010).

\bibitem{toy} R.~W.~Spekkens, Phys. Rev. A \textbf{75}, 032110, (2007).

\bibitem{liouville} S.~D.~Bartlett, T.~Rudolph and R.~W.~Spekkens, arXiv:1111.5057v1 [quant-ph].

\end{thebibliography}
\end{document}